\documentstyle[12pt,epsfig]{article}

\newcommand{\be}{\begin{equation}}
\newcommand{\ee}{\end{equation}}
\begin{document}  
\topmargin 0pt
\oddsidemargin=-0.4truecm
\evensidemargin=-0.4truecm
\renewcommand{\thefootnote}{\fnsymbol{footnote}}
\newpage
\vspace*{-2.0cm}  
\begin{flushright}
FISIST/00-00/CFIF\\
hep-ph/0012059
\end{flushright}
\vspace*{0.5cm}
\begin{center}
{\Large \bf Neutrino Magnetic Moment Solution for the Solar Neutrino Problem and the
SNO Experiment} 
\end{center}
\vspace{0.5cm}

\begin{center}
{\large 
Jo\~{a}o Pulido
\footnote{E-mail: pulido@beta.ist.utl.pt}\\  
\vspace{0.05cm}
{\em Centro de F\'\i sica das Interac\c c\~oes Fundamentais (CFIF)} \\
{\em Departamento de Fisica, Instituto Superior T\'ecnico }\\
{\em Av. Rovisco Pais, P-1049-001 Lisboa, Portugal}}\\
\vglue 0.8truecm

\begin{abstract}
The motivations for the magnetic moment solution to the solar neutrino problem are
briefly reviewed and the expected values for a number of observables to be measured
by the SNO experiment are calculated assuming three different solar magnetic field 
profiles. The observables examined are the charged current event rate, the ratio of 
the neutral current to the charged current event rates and the charged current 
electron spectrum as
well as their first and second moments. The dependence of results on the {\it{hep}}
neutrino flux is also analysed and a comparison is made with the corresponding 
oscillation results. 
\end{abstract}
\end{center}


The solar neutrino deficit may be explained on the assumption that neutrinos have a 
large magnetic moment $\mu_{\nu}=O(10^{-12})\mu_B$ which upon interaction with a 
magnetic field in the sun not exceeding $3\times10^{5}G$ may cause a flip in their helicity.
For Dirac and Majorana neutrinos we have respectively

\be
\nu_{{e}_{L}}\rightarrow\nu_{{e}_{R}}
\ee
\be
\nu_{{e}_{L}}\rightarrow\bar\nu_{\mu(\tau)_{R}}.
\ee
   
In the first case they become sterile for weak interactions while in the second 
they remain active for neutral currents, undergoing a simultaneous flip in
chirality and flavour \cite{SV}. This conversion process can be resonantly
enhanced in matter \cite{LMA}, which results in an energy dependent neutrino
deficit. Those neutrinos whose 
resonances are located in regions where the magnetic field is the largest will
undergo maximum suppression. Resonance location  
is determined by the neutrino energy, so that the smaller the energy, 
the higher the resonance density. In this way the resonance spin flavour 
precession of solar neutrinos in the matter and magnetic field of the sun 
(RSFP) implies an energy dependent neutrino deficit and possibly a most valuable 
information on the profile of the inner magnetic field of the sun.

RSFP has been overlooked by the physics comunity as a solution to the solar neutrino 
problem. The main reasons for this fact may be that it requires a large neutrino 
magnetic moment, several orders of magnitude above the electroweak standard model
prediction \cite{FS} and that no anticorrelation with the 11 year solar cycle seems 
to be observed in the data. This was the idea that a maximum of sunspot activity 
would be correlated with a minimum of the observed neutrino flux. However such 
would only be the case if resonances were located in the sunspot range (the outer
convective zone). Instead they could be located anywhere inside the sun, the depth of
sunspots is unknown and the field intensity is too small in sunspots to allow for a 
significant conversion. 

All oscillation scenarios including the two most favoured ones (LMA, LOW) \cite{BKS}
along with RSFP \cite{PA} indicate a sudden drop in the survival probability from the 
low energy (pp) to the intermediate energy neutrino sector ($^7$Be, CNO, pep) and 
a subsequent moderate rise as the energy increases further ($^8$B). RSFP offers a unique 
explanation for this probability shape, relating it to a sharp rise of the transverse 
magnetic field intensity somewhere in the solar interior, followed by a progressive, 
moderate decrease. Such a sharp rise is known to occur along the tachoclyne,
the region extending from the upper radiation zone to the lower convective zone where
the gradient of the angular velocity of solar matter is different from zero 
\cite{Parker}, \cite{A&A}. 
Two recent investigations \cite{GN}, \cite{PA} of a large number of field profiles have 
shown that the rate fits that can be obtained are excellent and far better 
than those for oscillations \cite{GC}, $\chi^2_{rates}(RSFP)<<\chi^2_{rates}(oscillations)$. 
In the present discussion the following three profiles will be considered
\be
B=0,~~~~~~x<x_{R}
\ee
\be
B=B_{0}\frac{x-x_{R}}{x_{C}-x_{R}},~~x_{R}\leq x\leq x_{C}
\ee
\be
B=B_{0}\left[1-\frac{x-x_{C}}{1-x_{c}}\right],~~x_{C}<x\leq 1
\ee
with $x_{R}=0.65$, $x_{C}=0.80$,
\be
B=0,~~~~~~x<x_{R}
\ee
\be
B=\frac{B_0}{cosh30(x-x_{R})},~~x\geq x_{R}
\ee
with $x_{R}=0.71$, and
\be
B=2.16\times10^{3},~~~~~~x\leq 0.7105
\ee
\be
B=B_{1}\left[1-\left(\frac{x-0.75}{0.04}\right)^2\right],~~0.7105<x<0.7483
\ee
\be
B=\frac{B_0}{cosh30(x-0.7483)},~~0.7483\leq x\leq 1
\ee
with $B_0=0.998B_{1}$ in the last two equations. The rate best fits were 
obtained respectively at 
$\Delta m^2_{21}=1.2\times10^{-8}eV^2$ and $B_{0}=1.23\times10^5G$,
$\Delta m^2_{21}=2.1\times10^{-8}eV^2$ and $B_{0}=1.45\times10^5G$ and
$\Delta m^2_{21}=1.6\times10^{-8}eV^2$ and $B_{0}=9.6\times10^4G$.

\begin{table}
{\small Table I: Double ratio $\bar{r}_{NC}=r_{NC}/r_{CC}$.
The values of the electron threshold energy for the CC reaction and
of the $hep$ neutrino flux scaling factor $f_{hep}$
are indicated in the parentheses in the first line.
The errors correspond to 90\% CL. }
\begin{center}
\begin{tabular}{ccccc} \hline \hline
Profile & $\bar{r}_{NC}$~(5 MeV; 1) &
$\bar{r}_{NC}$~(8 MeV; 1) & $\bar{r}_{NC}$~(5 MeV; 20) & $\bar{r}_{NC}$~(8
MeV; 20) \\
\hline
1 & $2.375\pm^{0.204}_{0.576}$ & \vspace*{0.2cm}
$2.320 \pm ^{0.200}_{0.566}$ &
$2.315\pm^{0.184}_{0.533}$ &
$2.222\pm^{0.177}_{0.511}$ \\
2 & $2.463\pm^{0.198}_{0.331}$ & \vspace*{0.2cm}
$2.410\pm^{0.187}_{0.329}$ &
$2.387\pm^{0.174}_{0.303}$ &
$2.294\pm^{0.165}_{0.296}$ \\
3 & $2.410\pm^{0.199}_{0.320}$ &
$2.347\pm^{0.196}_{0.308}$ &
$2.336\pm^{0.180}_{0.296}$ &
$2.242\pm^{0.173}_{0.284}$  \\
\hline
\end{tabular}
\end{center}
\end{table}

The following five SNO observables were investigated\cite{AP}\\
(a) the charged current (CC) electron spectrum
\be
S_{i}=\frac{\int_{T_i}^{T_{i+1}}dT\int_{Q}^{\infty}dEf(E)P(E)\int_{0}^{E-Q}
dT^{'}\frac{d\sigma_{CC}}{dT{'}}(T{'})R(T,T^{'})}
{\int_{T_i}^{T_{i+1}}dT\int_{Q}^{E_{max}}dEf(E)\int_{0}^{E-Q}
dT^{'}\frac{d\sigma_{CC}}{dT{'}}(T{'})R(T,T^{'})},
\ee
with the differential event rate given by
\be
\frac{dR_{CC}}{dT}\!\!=\!\!\!\int_{Q}^{\infty}\!\!\!\!\!dEf\!(\!E)\!P(\!E)\!\!\!\int_{0}^{\!E\!-\!Q\!}
\!\!\!\!\!dT^{'}\!\frac{d\sigma_{CC}}{dT{'}}\!(T{'})\!R(T,\!T^{'}\!)
\ee
and with the corresponding quantity in the denominator for standard neutrinos
($P(E)\rightarrow1$).\\
(b) the ratio of CC event rate to the standard CC event rate
\be
r_{CC}=\frac{\int_{T_m}^{\infty}dT\frac{dR_{CC}}{dT}}
{\int_{T_m}^{\infty}dT\frac{dR_{CC}^{st}}{dT}}=\frac{R_{CC}}{R_{CC}^{st}}
\ee
(c), (d) the first and second moments of the recoil electron kinetic energy distribution
\be
<T>=\frac{1}{R_{CC}}\int_{T_m}^{\infty}T\frac{dR_{CC}}{dT}dT
\ee
\be
\sigma=\sqrt{<T^2>-<T>^2}
\ee
(e) the double ratio of the non-standard to standard neutral current to charged current
event rates
\be
\bar{r}_{NC}=\frac{r_{NC}}{r_{CC}}=\frac{R_{NC}/R_{NC}^{st}}{R_{CC}/
R_{CC}^{st}}.
\ee 

In order to correctly incorporate the energy resolution function $R(T,T^{'})$, the
differential event rate per unit recoil electron energy must in fact be considered in
eqs. (11), (12) \cite{PA}. On the other hand for the neutral current event rate, one has in (16)
\be
R_{_{NC}}\!\!=\!\!\int_{\!E_B\!}^{\!\infty\!}\!\!\!f(\!E)[\!P(\!E)\sigma_{_{\!NC\!}}^{\nu d}
(\!E)\!+\!(1\!-\!P(\!E)) \sigma_{_{\!NC\!}}^{\bar{\nu}d}(\!E)]
\epsilon(\!E)d\!E.
\ee
Here the value $\epsilon(E)=0.5$ was used for the detector efficiency.

The observable $\bar{r}_{NC}$ is the best indicator for non-standard neutrino properties,
as the uncertainties inherent to fluxes and cross sections mostly cancel out \cite{AP}. 
Its predictions for 5 MeV and 8 MeV thresholds of the recoil electron kinetic energy
$T_m$ are given in Table I for a $hep$ neutrino flux equal to its BP'98 \cite{BP'98} solar model
prediction ($f_{hep}=1$) and ($f_{hep}=20$). A comparison between the corresponding 
oscillation predictions 
and RSFP ones can be seen from Tables I and II. It is seen that on the basis of the SNO 
experiment alone it will be hard to distinguish RSFP from other oscillation scenarios.
Assuming a peak magnetic field of the order of $3\times10^5G$ \cite{Parker}, \cite{A&A},
it is found that the neutrino magnetic moment should lie in the interval 
$\mu_{\nu}=(2-4)\times10^{-12}{\mu_B}$, consistent with most astrophysical bounds
\cite{magmo}.

\begin{table}
{\small Table II: Double ratio $\bar{r}_{NC}=r_{NC}/r_{CC}$ extracted
from ref. \cite{BKS} at 90\% CL (left column) and its possible
overlap with RSFP predictions given in Table I.}
\begin{center}
\begin{tabular}{ccc} \hline \hline
Osc. scenario & $\bar{r_{NC}}(5~MeV)$ & overlap with RSFP \\ 
              &                       &(90\%CL+90\%CL) \\
\hline
LMA & $3.37\pm^{0.89}_{0.55}$ & no \\
SMA & $2.53\pm^{0.79}_{0.65}$ & yes \\
LOW & $2.71\pm^{0.34}_{0.21}$ & yes \\
$\rm{VAC_S}$ & $2.67\pm^{0.98}_{0.65}$ & yes \\ 
$\rm{VAC_L}$ & $1.90\pm^{0.13}_{0.19}$ & no \\
Sterile & $0.96\pm^{0.02}_{0.02}$ & no \\ 
\hline
\end{tabular}
\end{center}
\end{table}


\begin{thebibliography}{pppp}
\bibitem{SV}  J. Schechter and J. W. F. Valle, Phys. Rev. {\bf D24}(1981) 1883, 
Erratum-{\it ibid.} {\bf D25} (1982) 283. 
\bibitem{LMA} C. S. Lim and W. J. Marciano, Phys. Rev. {\bf{D37}}, 1368 (1988); 
E. Kh. Akhmedov, Sov. J. Nucl. Phys. {\bf{48}}, 382 (1988); 
E. Kh. Akhmedov, Phys. Lett. {\bf{B 213}}, 64 (1988). 
\bibitem{FS} K. Fujikawa and R. E. Schrock, Phys. Rev. Lett. {\bf45} 963 (1980).
\bibitem{BKS} J. N. Bahcall, P. I. Krastev and A. Yu. Smirnov, Phys. Rev. {\bf62},
093004 (2000).
\bibitem{PA} J. Pulido and E. Kh. Akhmedov, Astrop. Phys. {\bf13}, 227 (2000).
\bibitem{Parker} E. N. Parker in "The Structure of the Sun", Proc. of the VI
Canary Islands School, Ed. Roca Cortes and F. Sanchez, Cambridge University Press
1996, p. 299.
\bibitem{A&A} H. M. Antia, S. M. Chitre, M. J. Thompson, astro-ph/0005587, to appear
in Astron. and Astrophys..
\bibitem{GN} M. Guzzo and H. Nunokawa, Astrop. Phys. {\bf12} 87 (1999).
\bibitem{GC} M. C. Gonzalez-Garcia, P. C. de Holanda, C. Pena-Garay. J. W. F. Valle,
Nucl. Phys. {\bf B573} 3 (2000).
\bibitem{AP} E. Kh. Akhmedov and J. Pulido, Phys. Lett. {\bf B485} 178 (2000).
\bibitem{BP'98} J. N. Bahcall, S. Basu and M. H. Pinsonneault, Phys. Lett. {\bf B433}
1 (1998).
\bibitem{magmo} S. I. Blinnikov, Institute for Theoretical and Experimental
Physics Report No. ITEP-88-19 (1988), unpublished; S. I. Blinnikov,
V.S. Imshennik, D.K. Nadyozhin, Sov. Sci. Rev. {\bf E} Astrophys. Space Sci. {\bf
6}, 185 (1987); G.G. Raffelt, Phys. Rev. Lett. {\bf 64} (1990) 2856; Astrophys. J.
{\bf 365} (1990) 559; 
V. Castellani and S. Degl'Innocenti, Astrophys. J. {\bf{402}}, 574 (1993). 

\end{thebibliography}
\end{document}